\documentclass[prd, amsfonts, twocolumn, nofootinbib, showpacs]{revtex4}
\usepackage{graphicx, epsfig}
\usepackage{color}
\usepackage{amsmath}
\usepackage{amssymb}
\newcommand{\be}{\begin{equation}}
\newcommand{\ee}{\end{equation}}
\newcommand{\bea}{\begin{eqnarray}}
\newcommand{\eea}{\end{eqnarray}}

\newcommand{\gapp}{\mathrel{\raise.3ex\hbox{$>$}\mkern-14mu
\lower0.6ex\hbox{$\sim$}}}
\newcommand{\lapp}{\mathrel{\raise.3ex\hbox{$<$}\mkern-14mu
\lower0.6ex\hbox{$\sim$}}}
\def\bbox{{\,\lower0.9pt\vbox{\hrule \hbox{\vrule height 0.2 cm
\hskip 0.2 cm \vrule  height 0.2 cm}\hrule}\,}}

\begin{document}
\title{ Serious limitations of the strong equivalence principle}
\author{De-Chang Dai}
\affiliation{ Institute of Natural Sciences, Shanghai Key Lab for Particle Physics and Cosmology, \\
and Center for Astrophysics and Astronomy, Department of Physics and Astronomy,\\
Shanghai Jiao Tong University, Shanghai 200240, China}

 %%%%%%%%%%%%%%%%%%%%%%%%%%%%%%%%%%%%%%%%%%%%%%%%%%%%%%%

\begin{abstract}
\widetext

 It is well known that an accelerated charged particle radiates away energy. However, whether an accelerated neutral composite particle radiates away energy is unclear. We study decoherent Larmor radiation from an accelerated neutral composite object. We find that the neutral object's long wavelength radiation is highly suppressed because  radiation from different charges is canceled out. However, the neutral object radiates high energy or short wavelength radiation without any suppression.
 In that case, radiation from each particle can be treated independently, and it is called the decoherent  radiation.   We compare a hydrogen atom's decoherent Larmor radiation with its gravitational radiation while the atom is in a circular orbit around a star. Gravitational radiation is stronger than the electromagnetic radiation if the orbital radius is larger than some critical radius.
 Since the decoherent radiation is related to the object's structure, this implies that the strong equivalence principle which states that gravitational motion does not depend on an object's constitution has severe limitations.

\end{abstract}

%%%%%%%%%%%%%%%%%%%%%%%%%%%%%%%%%%%%%%%%%%%%%%%%%%

\maketitle

\section{introduction}

 Since Galileo discovered that objects of any weight and composition fall toward the Earth at the same rate, the equivalence principle has played an important role in most of the theories of gravity. Einstein formulated the General Relativity based on the acceleration equivalence principle. As a consequence, a phenomenon observed in an accelerated system might also be observed in a gravitational field, e.g. Hawking radiation and Unruh radiation\cite{Unruh:1976db,Hartle:1976tp}.

There is no doubt that gravity and acceleration share common physical features. But Galileo's original observation that objects of any weight and composition respond to gravity equally is still subject to questioning. It is known that an accelerated charged particle radiates away energy through the Larmor radiation. This implies that a charged particle will emit radiation under gravitational acceleration and its trajectory is not going to be the same as a neutral particle's. The charged particle's equation of motion satisfies  Abraham-Lorentz-Dirac equation in curved spacetime instead of the simple geodesic path \cite{Abraham,Lorentz,Dirac,DeWitt}.

 Since the free falling objects' trajectories depend on their charges, Galileo's original idea must be modified. However, most of the falling objects are not charged.  So far only accelerated charged particles were studied in this context \cite{1966AmJPh..34Q1206R,1980AnPhy.124..169B}. However, if the composite particle is neutral,  the individual charges cancel out. It is still not clear whether a neutral composite object follows a trajectory different from the geodesic path.  It is also unknown whether a neutral composite object radiates away energy, since the radiation might allow an observer to reconstruct the distribution of matter which made a black hole\cite{Dai1,Dai:2016sls}.

To study this question, we first note that a charged particle is different from a neutral particle, because it radiates away electromagnetic fields. Therefore if one can study radiation, then he could reconstruct the distribution of matter in a composite object.  Similar problem has been studied in the synchrotron radiation theoretically and experimentally\cite{Nodvick(1954),Hirschmugl,Carr:2002mz,Orlandi(2002),Orlandi(2006),Hosaka et al.(2013)}. In general, electrons are accelerated in groups in a synchrotron. It is found that each electron radiates away electromagnetic field as a single charged particle if the radiation wavelength is shorter than the distance between electrons in the electron bundle. This is called decoherent radiation. Charged particles in that case are independent from each other and must be treated independently. However, if the electromagnetic field's wavelength is longer than the size of the electron bundle, the radiation is the same as a single charged particle with charge $Ne$ ($N$ is the number of electrons). The radiation is highly enhanced and is called coherent radiation.

Based on the coherent radiation theory, radiation from a neutral composite object is highly suppressed for long wavelength mode, because the object must be treated as a coherent single object. But the object can radiate away short wavelength modes, because these modes are radiated by all the components independently. We study this coherent radiation and decoherent radiation with the use of an interference function $f(\omega)$. If $f(\omega) \rightarrow 0$, the radiation is coherent. We find that whether the radiation is coherent or decoherent depends on the object's size. For a charged particle pair with distance $l$, the critical energy is $\omega_c\approx 1/l$. For a hydrogen atom (its radius is about $a_0$) the critical energy is $\omega_c\approx 1/a_0$.  We compare an accelerated hydrogen atom's electromagnetic radiation to its gravitational radiation,  while the atom is in a circular orbit around a star with mass equal to $M_\odot$, and find that the gravitational radiation is stronger than the electromagnetic  radiation if its orbital radius is larger than some critical radius.  In the following we introduce the coherent radiation theory and use a charged particle pair as a simple example. Then we calculate the hydrogen atom's interference function.

\section{Coherent radiation from a multi-particle system}

It is known that an accelerated charged particle radiates away energy, which is called Larmor radiation. Its intensity distribution is \cite{Jackson}

\begin{equation}
\frac{d^2I}{d\omega d\Omega}=\frac{\omega ^2 e^2}{16\pi^3 } \Bigr| \int\hat{n}\times(\hat{n}\times \vec{\beta})e^{i\omega(t-\hat{n}\cdot \vec{r})}dt\Bigr|^2
\end{equation}

Here $\vec{\beta}$ is the velocity of light ($\vec{v}/c$), $\hat{n}$ is the direction from the charged particle to an observer at infinity, $\vec{r}$ is the location of the particle. $\omega$ is the energy of the particle, $e$ is the particle's charge, and $\Omega$ is the solid angle. We set $c=\hbar=1$. In general, an object is made of several particles which have their own charges. The radiation comes from each charged particle

\begin{equation}
\label{rad1}
\frac{d^2I}{d\omega d\Omega}=\frac{\omega ^2 e^2}{16\pi^3 } \Bigr| \int\hat{n}\times(\hat{n}\times \vec{J})e^{i\omega(t-\hat{n}\cdot \vec{r})}dtdx^3\Bigr|^2
\end{equation}
$\vec{J}=\rho \vec{\beta}$, $\rho$ is the charge density. Each particle's radiation interferes with radiation from other partcles. The object's center of mass is at $\vec{r}_c$. $\vec{r}$ can be transfered to the center of mass frame, $\vec{r'}$, with the relation $\vec{r}=\vec{r}_c-\vec{r'}$. Then the radiation can be written as\cite{Hirschmugl}

\begin{eqnarray}
\label{rad2}
\frac{d^2I}{d\omega d\Omega}&=&\frac{\omega ^2 e^2}{16\pi^3 }f(\omega) \Bigr| \int \hat{n}\times(\hat{n}\times \vec{\beta})e^{i\omega(t-\hat{n}\cdot \vec{r}_c)}dt \Bigr|^2\\
\label{inter}
f(\omega)&=& \Bigr| \int \rho e^{i\omega\hat{n}\cdot \vec{r'}}dx'^3\Bigr|^2
\end{eqnarray}

Equation \ref{rad2} is similar to a single particle Larmor radiation multiplied by an interference factor $f(\omega)$. Therefore, if $f(\omega)$ is obtained, then the radiation is known. In the following, we will focus on calculating $f(\omega)$.

\section{A charged particle pair}

\begin{figure}
   \centering
\includegraphics[width=4cm]{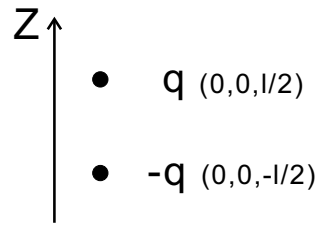}
\caption{There are two charged particles in the system. One particle with charge $q$ is at $(0,0,l/2)$ and the other particle with charge $-q$ is at $(0,0,-l/2)$.}
\label{two}
\end{figure}

The simplest composite object is made of a charged particle pair.  We assume that there is one particle with charge $q$ located at $(0,0,l/2)$, and another particle with charge $-q$ located at $(0,0,-l/2)$ in the center of mass frame (as in fig. \ref{two}). The center of mass is located at $\vec{r}_c$.  While in general particles may be free to move with respect to each other, we calculate only instantaneous radiation and keep the structure fixed. The charge density can be written with two delta functions,

\begin{equation}
\rho=\delta(\vec{r}-\frac{l}{2}\hat{z})-\delta(\vec{r}+\frac{l}{2}\hat{z}).
\end{equation}

To simplify the study, we consider non-relativistic case, and the Lorentz contraction is neglected. The interference function $f(\omega)$ is
\begin{equation}
f(\omega)=\Bigr| e^{i\omega \frac{l}{2}\hat{n}\cdot \hat{z}}-e^{-i\omega \frac{l}{2}\hat{n}\cdot \hat{z}}\Bigr|^2=4\sin^2(\omega \frac{l}{2}\hat{n}\cdot \hat{z}).
\end{equation}

\begin{figure}
   \centering
\includegraphics[width=6cm]{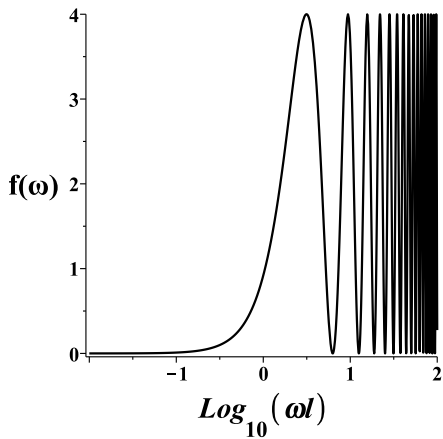}
\caption{The interference function for two charged particles, $f(\omega)$, as a function of $\omega$. For low $\omega$, radiation from these two particles is coherent and cancel out each other. The observer is in the direction of the  z-axis. For high $\omega$ modes, the interference pattern appears. It oscillates between $0$ and $4$ and the oscillation period is $\Delta\omega=\frac{4\pi}{l}$. On average, it becomes equal to $2$. For a finite size detector, one will not be able to distinguish the oscillations and it is as if the radiation is coming from two independent charged particles.}
\label{along-z}
\end{figure}

This is the same as double slit interference with phase difference $\pi$. If the observer is at $(0,0,\infty)$,  modes with $\omega \ll 1/l$ cancel out each other  and $f(\omega)\rightarrow 0$. The interference pattern appears for $\omega l\gg 1$ (fig. \ref{along-z}). The intensity of the interference is between $0$ and $4$. The oscillation cycle is $\Delta\omega=\frac{4\pi}{l}$. While $\omega$ is large enough, a detector will not be able to pick up the oscillation pattern and it will record only the average effect. On average, the intensity is $\bar{f}(\omega)=2$. This value is what one expects from two independent charged particles. Therefore even though the object's total charge is $0$, the high energy radiation is the same as the radiation from two independent particles.

\begin{figure}
   \centering
\includegraphics[width=6cm]{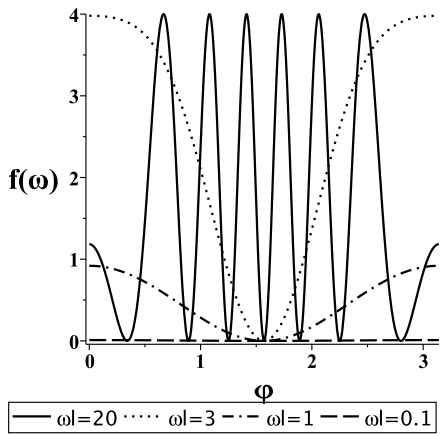}
\caption{There are four different interference functions, $f(\omega)$, as a function of $\omega l$ in the figure. The x-axis is the angle between the observer's direction and z-axis, $\varphi=\cos^{-1}(\hat{n}\cdot \hat{z})$ . For $\omega l\ll 1$, $f(\omega)$ is almost $0$, because radiation from these two charged particles cancel out. For $\omega l\gg 1$, the interference pattern starts to appear and its period becomes shorter and shorter with respect to $\varphi$. When a detector cannot distinguish the oscillation, the radiation is identical to radiation from two independent particles. }
\label{along-theta}
\end{figure}

For an observer at different location, the interference can be described by the angle $\varphi = \cos^{-1}(\hat{n}\cdot \hat{z})$. Again the interference function, $f(\omega)$, is canceled out for $\omega \ll 1/l$ and the interference pattern appears while $\omega \gtrapprox 1/l$(fig. \ref{along-theta}). The oscillation pattern is between $0$ and $4$ and oscillation changes much quicker for larger $\omega$. Its average is $\bar{f}(\omega)=2$. This proves that the radiation with shorter wavelength acts as if it is coming from two independent charged particles, or in the other word two charged particles radiate away energy decoherently.

\section{hydrogen}

A neutral atom has no net charge. At first glimpse it seems that an accelerated atom will not radiate any electromagnetic wave. However, since an atom has its own structure, its high $\omega$ radiation should not canceled out. This can be tested by calculating the interference function $f(\omega)$. To make it simple, we will focus on the simplest atom, hydrogen. A hydrogen has a proton in the center and an electron circulating around it. The acceleration will distort the hydrogen's wavefunction \cite{Dai2}. Here we focus on small acceleration and assume the distortion can be neglected. We also consider low velocities so that the Lorentz distortion is neglected.  The electron's wave function in the center of mass frame is

\begin{eqnarray}
&&\phi^0_{n,l,m}(r,\theta,\phi)=R_{n,l}Y^m_l(\theta,\phi)\\
&&R_{n,l}\sim\exp(-\frac{r}{na_0})\Big(\frac{r}{na_0}\Big)^l L^{2l+1}_{n-l-1}(\frac{2r}{na_0})
\end{eqnarray}

$L^{2l+1}_{n-l-1}(x)$ is the associated Laguerre polynomial. $Y^m_l(\theta,\phi)$ is the spherical harmonic function of degree $l$ and order $m$. $a_0$ is the Bohr radius, $a_0=\frac{1}{me^2}=5.2917721067\times 10^{-11}$m. The ground and first excited states' wavefunctions are

\begin{eqnarray}
\phi^0_{1,0,0}&=&\Big(\frac{1}{\pi a_0^3}\Big)^{\frac{1}{2}}\exp(-\frac{r}{a_0})\\
\phi^0_{2,0,0}&=&\Big(\frac{1}{32\pi a_0^3}\Big)^{\frac{1}{2}}\Big(2-\frac{r}{a_0}\Big)\exp(-\frac{r}{2a_0})\\
\phi^0_{2,1,0}&=&\Big(\frac{1}{32\pi a_0^3}\Big)^{\frac{1}{2}}\frac{r}{a_0}\exp(-\frac{r}{2a_0})\cos\theta\\
\phi^0_{2,1,\pm 1}&=&\mp\Big(\frac{1}{64\pi a_0^3}\Big)^{\frac{1}{2}}\frac{r}{a_0}\exp(-\frac{r}{2a_0})\sin\theta e^{\pm i\phi}
\end{eqnarray}

The electron's probability function is obtained from its wave function

\begin{eqnarray}
\rho_e&=&\phi^*\phi
\end{eqnarray}
, and the proton can be treated as a point particle at the center.
\begin{equation}
\rho_p=\delta (\vec{r})
\end{equation}

An electron has a negative charge and a proton has a positive charge, so the total charge distribution is

\begin{eqnarray}
\rho&=&\rho_p-\rho_e
\end{eqnarray}

\begin{figure}
   \centering
\includegraphics[width=7cm]{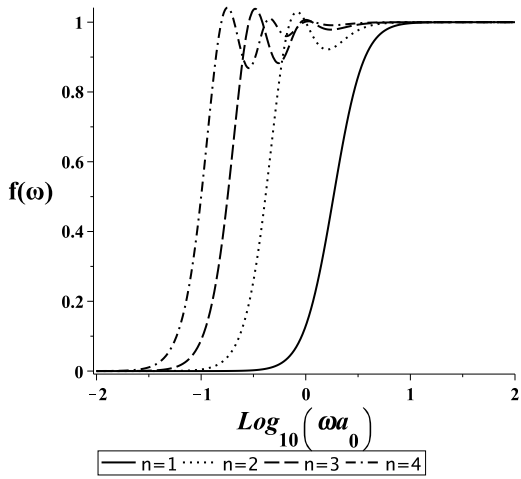}
\caption{The four curves are the interference functions for different quantum states, $(n,0,0)$. Since the larger $n$ state has a larger radius, radiation becomes decoherent for much smaller $\omega$. Since all of the 4 quantum states are spherically symmetric, the observer's direction does not affect the interference pattern.  }
\label{f-omega}
\end{figure}

The interference function can be calculated directly from equation \ref{inter}. Figure \ref{f-omega} shows the interference function of s-wave states( $l=0$, $m=0$). $f(\omega)$ is canceled out for small $\omega$. While $\omega$ increases,  $f(\omega)$ increases and achieves $1$ at very large $\omega$. The higher $n$ states become decoherent at lower $\omega$ than the lower $n$ states did, because the higher $n$ states have larger radii. However, the decoherent radiation is not a combination of the proton and electron. The electron's radiation is canceled out by itself, because of its own density distribution. Only the proton's radiation is left. The atom radiates high $\omega$ modes as if it is a single proton.

If the acceleration does not change very quick, there will not be much high frequency radiation ($\tilde{\beta}(\omega)\rightarrow 0$ as $\omega a_0\rightarrow \infty$ ). In that case we find the ground state's $f(\omega )$ is

\begin{eqnarray}
f(\omega)&=& \Bigr| \int \rho e^{i\omega\hat{n}\cdot \vec{r}_c}dx^3\Bigr|^2\nonumber\\
&=&\frac{\omega^4 a_0^4(\omega^2a_0^2+8)^2}{(16+8\omega^2 a_0^2+\omega^4 a_0^4)^2}
\end{eqnarray}

In a low $\omega$ approximation

\begin{equation}
f(\omega) \approx\frac{\omega^4 a_0^4}{4}
\end{equation}

We consider non-relativistic approximation. The velocity is much lower than the speed of light, $\beta\ll 1$.

\begin{eqnarray}
&&\Bigr|\int \hat{n}\times(\hat{n}\times \vec{\beta})e^{i\omega(t-\hat{n}\cdot \vec{r})}dt\Bigr|^2\nonumber\\
&&\approx \Bigr|\int \sin(\alpha)\beta e^{i\omega t}dt\Bigr|^2\nonumber\\
&&=\sin^2(\alpha) \tilde{\beta}^2
\end{eqnarray}

Here, $\tilde{\beta}$ is $\beta$'s Fourier transform,
\begin{equation}
\tilde{\beta}=\int \beta e^{i\omega t}dt.
\end{equation}

From equation \ref{rad2}, the intensity is

\begin{eqnarray}
\frac{d^2I}{d\omega d\Omega}&=&\frac{\omega ^2 e^2}{16\pi^3 }\frac{\omega^4 a_0^4}{4}\sin^2(\alpha) \tilde{\beta}^2
\end{eqnarray}

The total released energy is

\begin{eqnarray}
I&=&\int \frac{\omega ^2 e^2}{16\pi^3 }\frac{\omega^4 a_0^4}{4}\sin^2(\alpha) \tilde{\beta}^2d\omega d\Omega\nonumber\\
&=&\int_0^{\infty} \frac{e^2 \omega^6a_0^4}{24\pi^2}\tilde{\beta}^2 d\omega\nonumber\\
\label{T-rad}
&=&\int_{-\infty}^{\infty} \frac{e^2 a_0^4}{24\pi }(\partial_t^3\beta)^2 dt
\end{eqnarray}

This implies that the energy flux is $\frac{e^2 a_0^4}{24\pi }(\partial_t^3\beta)^2$, if the acceleration is  changing slowly.

\section{Comparing neutral atom's electromagnetic and gravitational radiation}

It is known that an accelerated particle also radiates away gravitational waves. A freely falling composite object may radiate both gravitational waves and electromagnetic waves. To simplify the case, we study a composite particle under the gravitational influence of a heavy object with mass $M$. We consider a circular orbit. Then the average power of gravitational radiation over one period of the  motion is \cite{Peters}

\begin{equation}
\label{radiation_gravity}
<\frac{dI_g}{dt}>=\frac{32G^4}{5}\frac{M^2m^2(M+m)}{R^5}.
\end{equation}

Here $R$ is the orbital radius and $m$ is the composite particle's mass. In this study we assume that the composite particle is a hydrogen atom in his ground state. The orbital angular frequency is $\omega_o^2 = \frac{GM}{R^3}$.

The composite particle's electromagnetic radiation can be found from equation \ref{T-rad},
\begin{equation}
\label{radiation_em}
\frac{dI}{dt}=\frac{e^2 a_0^4 R^2 \omega_o^8}{24\pi }=\frac{e^2 a_0^4 G^4M^4}{24\pi R^{10}}.
\end{equation}

One can compare equation \ref{radiation_gravity} with equation \ref{radiation_em}. The gravitational radiation is stronger than the  electromagnetic radiation if its orbital radius is larger than the critical radius,

\begin{equation}
R_c=\Bigg( \frac{5e^2a_0^4M}{768\pi m^2}\Bigg)^{1/5}
\end{equation}

If the heavy object is of the sun's mass ($M=1M_\odot$), the critical radius is $R_c=1.96 \text{m}$. This is much smaller than the sun's radius. Therefore, in the sun's case,  most a freely falling hydrogen atom still radiates away much more gravitational than electromagnetic energy. However, this may change for a very massive and compact object.

\section{Conclusion}

We studied the decoherent Larmor radiation from a neutral object which is made of charged particles. This radiation depends on the object's structure. For a charged particle pair separated by the distance $l$, the characteristic energy is $\omega_c \approx 1/l$. The object's low $\omega$ radiation ($\omega\ll \omega_c$) must be treated as if it is coming from a single object and radiation is highly suppressed. The object's high $\omega$ modes ($\omega\gg \omega_c$) must be treated as if they are coming from independent particles. They are radiated decoherently.

The same argument is applied to the hydrogen atom. A hydrogen atom's radius is about $a_0$. Its characteristic energy is $\omega_c\approx 1/a_0$. The $\omega\ll \omega_c $ modes are highly suppressed, because the charged particles radiate coherently. The $\omega\gg \omega_c$ modes are radiated decoherently and one can treat the electron and proton independently. However, at this  energy scale electrons cannot be treated as single independent  particles. Electron's radiation is canceled out because of electron's charge density distribution. The high $\omega$ modes are mainly from the proton. The radiation is similar to the radiation from  a single charged particle, instead of the  radiation from two charged particles. If the acceleration does not change very quickly, the high $\omega$ radiation is weak. In this case the total radiation depends on the  third time derivative of velocity (equation \ref{T-rad}).

An accelerated neutral atom emits not only electromagnetic waves but also gravitational waves. Both of these processes are highly suppressed. The electromagnetic radiation is suppressed by charge cancellation, while the gravitational radiation is suppressed by the weak gravitational coupling.  We compare these two types of radiations for a hydrogen atom orbiting a star of solar mass. We find the gravitational radiation is much stronger than the electromagnetic radiation if orbital radius is larger than  $R_c=1.96 \text{m}$.

This result shows that an accelerated particle will radiate energy according to its structure. This radiation may be highly suppressed, because of the constituent particles' interference, but it is not fully canceled out. Therefore if we drop two objects with different internal structures in a gravitational field, these two objects will radiate away different amounts of energy and will fall at a different rate. The strong equivalence principle which states that gravitational motion does not depend on an object's constitution must be revised. It is also possible to reconstruct the falling matter's internal structure through the radiation\cite{Dai1}.

\begin{acknowledgments}
D.C. Dai was supported by the National Science Foundation of China (Grant No. 11433001 and 11447601), National Basic Research Program of China (973 Program 2015CB857001), No.14ZR1423200 from the Office of Science and Technology in Shanghai Municipal Government, the key laboratory grant from the Office of Science and Technology in Shanghai Municipal Government (No. 11DZ2260700) and  the Program of Shanghai Academic/Technology Research Leader under Grant No. 16XD1401600.

\end{acknowledgments}

\end{document}